\documentclass{ws-ijmpa}
\usepackage[super,compress]{cite}
\usepackage{graphicx}
\begin{document}
\markboth{{\it J. C. Fabris, O. F. Piattella, D. C. Rodrigues, B. Chauvineau \& M. H. Daouda}}{{\it Introducing quantum effects in classical theories}}

%
\catchline{}{}{}{}{}
%

\title{Introducing quantum effects in classical theories}

\author{J. C. Fabris, O. F. Piattella, D. C. Rodrigues
}

\address{Departamento de F\'{\i}sica, Universidade Federal do Esp\'{i}rito Santo,\\
Avenida Fernando Ferrari, 514, Vit\'oria, Esp\'{\i}rito Santo, CEP29060-300, Brazil\\
{\it fabris@pq.cnpq.br}}

\author{B. Chauvineau}

\address{Laboratoire Lagrange (UMR 7293), Universit\'{e} de Nice-Sophia Antipolis, CNRS, \\Observatoire de la C\^{o}te d'Azur, BP 4229, 06304 Nice cedex 4, France\\
{\it Bertrand.Chauvineau@oca.eu}}

\author{M. H. Daouda
}

\address{D\'epartament de Physique - Universit\'e de Niamey, Niamey, Niger\\
{\it daoudah8@yahoo.fr}
}

\maketitle

\begin{history}
\received{Day Month Year}
\revised{Day Month Year}
\end{history}

\begin{abstract}
In this paper, we explore two different ways of implementing quantum effects in a classical structure. The first one is through an external field. The other one is modifying the classical conservation laws. In both cases, the consequences for the description of the evolution of the universe are discussed.

\keywords{Cosmology; external fields; Rastall's theory.}
\end{abstract}

\ccode{PACS numbers: 04.20.-q, 98.80.-k}


\section{Introduction}

The General Relativity Theory (GR) is, perhaps, the simplest and most attractive geometric theory of gravity already conceived. It comes out from the most economic variational principle for a geometric theory and
it leads to second order (non-linear) differential equations. As consequence, the Cauchy problem is well-posed, and it is free from intrinsic instabilities. From observational and experimental side, GR is in agreement with all local tests and it conducts to models for stars and universe that reproduces many observed features\cite{mukha}.
However, GR has also many drawbacks. It is plagued with singularities in many contexts. In cosmology, it needs two exotic components (dark matter and dark energy), representing $95\%$ of the total cosmic budget, to fit the observations at
large scales. These exotic components remain without any direct and experimental evidence.
At astrophysics level, the cosmological simulations indicate problems with the standard $\Lambda$CDM model, like excess of substructures around galaxies, divergence in the density in the central region of galaxies, among others \cite{wei}.

In view of these problems, many alternatives have been proposed to replace General Relativity \cite{amendola}. A very popular alternative today are the
$f(R)$ theories\cite{felice}. It is a natural generalisation of the Einstein-Hilbert Lagrangian, but care must be taken considering stabilities, since it is a higher derivative gravity.
Another possibility is the Galileons theories, based on scalar fields with a special translational symmetry \cite{galleon}.
Connected somehow with the Galileons theories, there is the Horndeski theory which is the most general combination of derivative of scalar fields, leading to second order differential equations \cite{horndeski}.

All these generalisations face a huge problem: While they are able to explain the behaviour of the universe in large scales
without a dark sector, they fail to reproduce the local tests of gravity for the same range of values for the characteristic parameters of the theory.
In order to cope this problem, a {\it screening mechanism} is conceived which blocks the propagation of the new degrees of freedom (the fifth force in occurrence) at local level.

There are different types of screening mechanisms: Chameleon, for which the mass of the new field depends on the mean density of the environment;
Symmetron which implies that the coupling of new field depends on the density of the environment;
Vainshtein mechanism, for which the kinetic term becomes large in high density environment, suppressing the coupling to matter.
All these mechanisms may lead (in principle!) to viable models. For a revision on these different mechanisms, see for example reference \cite{koyama}.

In what follows, we explore two other different possibilities, the external field approach and the modification of the classical conservation laws. Both cases may be viewed, in some sense, as a classical way to introduce quantum effects. The connection to quantum effects is kept in a rather phenomenological level. But, 
it is expected that at least some important features of the possible counter-reaction in the classical equations due to those quantum effects are retained in both
proposals we will describe below.

\section{Alternative Theories of Gravity: The Brans-Dicke Theory}

One of the first alternative to General Relativity was the Brans-Dicke theory, which tried to implement consistently the possible variation of the gravitational coupling (and the Mach's principle) through a long-range scalar field
non-minimaly coupled to gravity.
Its Lagrangian reads \cite{bd}:
\begin{eqnarray}
{\cal L} = \sqrt{-g}\biggr\{\phi R - \omega\frac{\phi_{;\rho}\phi^{;\rho}}{\phi}\biggl\} + {\cal L}_m,
\end{eqnarray}
where $\omega$ is a new coupling parameter and ${\cal L}_m$ is the matter Lagrangian. When $\omega \rightarrow \infty$, the General Relativity Theory is recovered (or almost, see for example reference \cite{romero}).

The corresponding field equations are:
\begin{eqnarray}
R_{\mu\nu} - \frac{1}{2}g_{\mu\nu}R &=& \frac{8\pi}{\phi}T_{\mu\nu} + \frac{\omega}{\phi^2}(\phi_{;\mu}\phi_{;\nu} - \frac{1}{2}g_{\mu\nu}\phi_{;\rho}\phi^{;\rho})\nonumber\\
&+& \frac{1}{\phi}(\phi_{;\mu;\nu} - g_{\mu\nu}\Box\phi),\\
\Box\phi &=& \frac{8\pi}{3 + 2\omega}T,\\
{T^{\mu\nu}}_{;\mu} &=& 0.
\end{eqnarray}

One of the problems with the Brans-Dicke theory is that, in order to satisfy the local tests, some estimations give $\omega > 40.000$ \cite{will}.
Hence, the theory is essentially indistinguishable from General Relativity. But, we must remark that in this original formulation, the Brans-Dicke theory has no potential.  If we introduce a potential and perform a conformal transformation in order to obtain a minimal coupling formulation of the theory, we have the 
following equations for the scalar field and for the matter term:
\begin{eqnarray}
\Box\sigma &=& - V_\sigma - \kappa Qe^{\sigma}\bar T = - V_\sigma^{eff}\\
{{\bar T}^{\mu\nu}}_{;\mu} &=& Q\bar T e^\sigma\sigma^{;\nu},
\end{eqnarray}
with $\sigma = \ln\phi$, and where $\kappa$ is the gravitational coupling and $Q$ a parameter connected with the conformal transformation.
It is important to remark that the effective potential $V^{eff}$ depends on the density. This fact may inhibit the propagation of the scalar degree of freedom in high density environment, reducing the theory essentially to GR, but allowing it in low density environment. Hence, the departure from GR becomes relevant only at cosmological level.
This is somehow on the basis of the Chameleon mechanism.

\section{External Field}

We can define an {\it external field} as a field not subjected to the variational principle.
In the process of deriving the field equations from an action, the variation with respect to all the fields are considered, except for the external ones.
In particular, this procedure is justified if the expression of the external field as a space-time function is known beforehand. It is also called sometimes as a {\it non-dynamical field}.
One of the relevance of the external field approach is the fact that it may be a procedure to include in a given system defined by a Lagrangian the influence of external factors. It is somehow connected with the principle of {\it open systems}.
One simple exemple is the problem of a trajectory of a charge submitted by a constant, fixed, electric field.

We will present a specific exemple of how to implement the idea of external field based on reference \cite{chauvineau}.
Let us consider now the Brans-Dicke action, but with the scalar field representing an external field. Moreover, we consider a cosmological
term depending on this scalar (external) field.
\begin{eqnarray}
{\cal L} = \sqrt{-g}\biggr\{\phi R - \omega(\phi)\frac{\phi_{;\rho}\phi^{;\rho}}{\phi }- 2\phi\Lambda(\phi)\biggl\} + {\cal L}_m.
\end{eqnarray}
It has been supposed also that the Brans-Dicke coupling may depend on the scalar field.
The field equations are now:
\begin{eqnarray}
R_{\mu\nu} - \frac{1}{2}g_{\mu\nu}R + g_{\mu\nu}\Lambda &=& \frac{8\pi}{\phi}T_{\mu\nu} + \frac{\omega}{\phi^2}(\phi_{;\mu}\phi_{;\nu} - \frac{1}{2}g_{\mu\nu}\phi_{;\rho}\phi^{;\rho})\nonumber\\
&+& \frac{1}{\phi}(\phi_{;\mu;\nu} - g_{\mu\nu}\Box\phi),\\
{T^{\mu\nu}}_{;\mu} &=& 0.
\end{eqnarray}
The conservation of the energy-momentum tensor is still valid because of the differmorphism: the matter Lagrangian does not depend on the external field
\cite{chauvineau}.

However, the Bianchi's identities are still valid, leading to
\begin{eqnarray}
\biggr\{\Box\phi - \frac{1}{3 + 2\omega}\biggr[8\pi T + \omega'\phi_{;\rho}\phi^{;\rho} + 2\phi(\Lambda - \phi\Lambda'\biggl]\biggl\}\phi_{;\mu} = 0,
\end{eqnarray}
where the primes mean derivative with respect to the scalar field $\phi$.
The very important difference with respect to the usual Brans-Dicke theory is that now, 
$\phi = \mbox{constant}$
is a solution for any $\omega$ and even in presence of matter. This is not possible in the Brans-Dicke theory.

We may apply this construction to cosmology. Let us consider the flat FLRW metric,
\begin{eqnarray}
ds^2 = dt^2 - a(t)^2(dx^2 + dy^2 + dz^2).
\end{eqnarray}
For simplicity, we ignore the cosmological term $\Lambda$ and we take $\omega$ as a constant.
Moreover, let us suppose a pressureless matter. The equations of motion are the following:
\begin{eqnarray}
3\biggr(\frac{\dot a}{a}\biggl)^2 &=& \frac{8\pi \rho}{\phi} + \frac{\omega}{2}\frac{\dot\phi^2}{\phi^2} - 3\frac{\dot a}{a}\frac{\dot\phi}{\phi},\\
2\frac{\ddot a}{a} + \biggr(\frac{\dot a}{a}\biggl)^2 &=& - \frac{\omega}{2}\frac{\dot\phi^2}{\phi^2} - 2\frac{\dot a}{a}\frac{\dot\phi}{\phi} - \frac{\ddot\phi}{\phi},\\
\bigg\{\ddot\phi + 3\frac{\dot a}{a}\dot\phi - \frac{8\pi\rho}{3 + 2\omega}\biggl\}\dot\phi &=& 0,\\
\dot\rho + 3\frac{\dot a}{a}\rho &=& 0.
\end{eqnarray}

The cosmological solution for General Relativity (flat case), in the matter dominated phase is given by:
\begin{eqnarray} 
a(t) = A(t - t_0)^{2/3},\quad \phi &=& C.
\end{eqnarray}
The general flat Brans-Dicke solutions for $p = 0$ are,
\begin{eqnarray} 
a(t) &=& B(t - t_+)^{q_+}(t - t_-)^{q_-},\\
\phi(t) &=& \phi_0(t - t_+)^{p_+}(t - t_-)^{p_-}.
\end{eqnarray}
with
\begin{eqnarray}
q_{\pm} &=& \frac{1 + \omega \pm \sqrt{1 + \frac{2}{3}\omega}}{4 + 3\omega},\\
p_{\pm} &=& \frac{1 \pm 3\sqrt{1 + \frac{2}{3}\omega}}{4 + 3\omega},
\end{eqnarray}
where $t_\pm$ are integration constants with $t_+ > t_-$.
In order to have positive energy and positive gravitational coupling we must have $ \omega > - \frac{4}{3}$
These Brans-Dicke solutions were described in reference \cite{gurevich}.

The GR and BD solutions can not be matched in the traditional context: They are two different theories.
The situation changes if $\phi$ is now an external field. 
Let us call such theory, {\it EST} for External Scalar Theory. In this case, the above described GR solution is also a solution for gravitation-external field system.

Let us fit the solutions in the external field approach. We fix $t_0 = 0$, and the transition time $t = t_m > t_+$.
The solutions for the matter component density in each phase read,
\begin{eqnarray}
6\pi \rho_{GR}\left( t\right)  &=&\frac{C}{t^{2}},
\label{dust densities} \\[.1in]
4\pi \rho _{BD}\left( t\right)  &=&\frac{3+2\omega }{4+3\omega }\frac{%
\phi _{0}}{\left( t-t_{+}\right) ^{3q_{+}}\left( t-t_{-}\right) ^{3q_{-}}}. 
\end{eqnarray}%
The subscripts $GR$ and $BD$ stand for the different phases, the GR phase and the BD phase, respectively.
The continuities of the scalar and its (logarithmic) derivative, of the
scale factor and its (logarithmic) derivative, give respectively
\begin{equation}
\phi _{0}\left( t_{m}-t_{+}\right) ^{p_{+}}\left( t_{m}-t_{-}\right)
^{p_{-}}=C,  \label{scalar cont}
\end{equation}
\begin{equation}
\frac{p_{+}}{t_{m}-t_{+}}+\frac{p_{-}}{t_{m}-t_{-}}=0,
\label{scalar-der cont}
\end{equation}
\begin{equation}
B\left( t_{m}-t_{+}\right) ^{q_{+}}\left( t_{m}-t_{-}\right)
^{q_{-}}=At_{m}^{2/3},  \label{scalefact cont}
\end{equation}
\begin{equation}
\frac{q_{+}}{t_{m}-t_{+}}+\frac{q_{-}}{t_{m}-t_{-}}=\frac{2}{3t_{m}}.
\label{scalefact der cont}
\end{equation}
The final expressions are the following:
\begin{eqnarray}
& &t_{\pm }=\frac{1}{1\pm 3\sqrt{1+\frac{2}{3}\omega }}t_{m},\\
& &B =A\left( 3\sqrt{1+\frac{2}{3}\omega }+1\right) ^{q_{+}}\left( 3\sqrt{1+%
\frac{2}{3}\omega }-1\right) ^{q_{-}}\nonumber\\
&\times&\left( 9+6\omega \right) ^{-\frac{%
1+\omega }{4+3\omega }}t_{m}^{\frac{2}{3\left( 4+3\omega \right) }},
\label{BD param Bphi} \\
& &\phi _{0} = C\left( 3\sqrt{1+\frac{2}{3}\omega }+1\right) ^{p_{+}}\left( 3%
\sqrt{1+\frac{2}{3}\omega }-1\right) ^{p_{-}}\nonumber\\
&\times&\left( 9+6\omega \right) ^{-%
\frac{1}{4+3\omega }}t_{m}^{-\frac{2}{4+3\omega }}.  
\end{eqnarray}

Now, it is possible to have a transition from a GR behaviour to a BD behaviour during a single
phase of the evolution of the universe. The success of the GR theory concerning the structure formation can be kept.
But, new features (accelerated expansion?) may be introduced non-trivially. Work is in progress in order to implement 
explicitly possible realistic scenarios using this structure. More details on these EST calculations, and extensions of the toy model, are available on reference\cite{chauvineau}.

\section{Rastall's Theory}

Quantum effects lead to an effective energy-momentum tensor
\begin{eqnarray}
 T^{\mu\nu} \quad \Rightarrow \quad{T^{\mu\nu}}_{eff},
 \end{eqnarray}
 which may differ substantially from the classical one \cite{birrell}.
 
 In 1972, P. Rastall \cite{rastall} has proposed a modification of the General Relativity. The new equations are:
\begin{eqnarray}
 R_{\mu\nu} - \frac{\lambda}{2}g_{\mu\nu}R &=& \kappa T_{\mu\nu},\\
{T^{\mu\nu}}_{;\mu} &=& \frac{1 - \lambda}{2\kappa}R^{;\nu}. 
\end{eqnarray}
The main argument to introduce such modification is the fact that the usual conservation laws are tested only in flat spacetime.
For $\lambda = 1$, General Relativity is recovered.

The Rastall's theory exploit the ambiguity in the conservation laws in General Relativity.
For a simple example of this ambiguity let us consider a fluid with an equation of state $p = \omega\rho$, $\omega = \mbox{constant}$. The conservation
equation is,
\begin{eqnarray}
\dot\rho &+& 3\frac{\dot a}{a}(1 + \omega)\rho = 0,\\
\rho &\propto& a^{-3(1 + \omega)}.
\end{eqnarray}
In spite of the expansion of the universe, the energy density remains constant for $\omega = - 1$ and even increases for
$\omega < - 1$. In some sense this is due to an exchange of energy of the fluid with gravity. However, the energy of gravity is
not clearly defined since GR is a geometric theory of gravitation \cite{wald}.

The parameter $\gamma$ can be seen as a deformation of the usual GR conservation laws. In this sense, the Rastall theory can be considered as a classical and phenomenological implementation of quantum effects.

The new equations can be re-written as,
\begin{eqnarray}
R_{\mu\nu} - \frac{1}{2}g_{\mu\nu}R &=& \kappa\biggr\{T_{\mu\nu} - \frac{\gamma - 1}{2}g_{\mu\nu}T\biggl\},\\
{T^{\mu\nu}}_{;\mu} &=& \frac{\gamma - 1}{2}T^{;\nu}.
\end{eqnarray}
Now, $\gamma = \frac{3\lambda - 2}{2\lambda - 1}.$
If $\lambda = 1$, $\gamma = 1$.
A fluid with zero pressure in the General Relativity context has, in Rastall's theory, the following effective equation of state:
\begin{eqnarray}
 \omega_{eff} = \frac{\gamma - 1}{3 - \gamma}.
\end{eqnarray}
Hence, a dust fluid can accelerate the universe.

We may explore the new possibilities for the description of the universe in the context of the Rastall's theory.
The present universe is very well described by the $\Lambda$CDM model. However, this description has some important drawbacks. Among then, there is the excess of power in the matter spectrum at
small scales, as already stated above. Can the Rastall's cosmology give some new point of view for these problems
preserving the success of the $\Lambda$CDM model?

Let us consider a two fluid model \cite{daouda}:
\begin{eqnarray}
R_{\mu\nu} - \frac{1}{2}g_{\mu\nu}R &=& 8\pi G\biggr\{T^m_{\mu\nu} + T^x_{\mu\nu} 
- \frac{\gamma - 1}{2}g_{\mu\nu}(T^m + T^x)\biggl\},\\
{T^{\mu\nu}_x}_{;\mu} &=& \frac{\gamma - 1}{2}(T_m + T_x)^{;\nu},\quad {T^{\mu\nu}_m}_{;\mu} = 0,
\end{eqnarray}
where $m$ and $x$ stand for the dark matter and dark energy component, respectively.
It is necessary to have the usual conservation law for one of the fluids in order to allow structure formation.

The equations of motion are the following:
\begin{eqnarray}
H^2 &=& \frac{8\pi G}{3}\biggr\{(3 - 2\gamma)\rho_x + \frac{-\gamma + 3}{2}\rho_m\biggl\},\\
\dot\rho_m + 3H\rho_m &=& 0,\nonumber\\
(3 - 2\gamma)\dot\rho_x &=& \frac{\gamma - 1}{2}\dot\rho_m,\\
\rho_m = \frac{\rho_{m0}}{a^3}&,&\quad \rho_x = \frac{\rho_{x0}}{3 - 2\gamma} + \frac{\gamma - 1}{2(3 - 2\gamma)}\rho_m. 
\end{eqnarray}

Combining these expressions we obtain,
\begin{eqnarray}
H^2 = \frac{8\pi G}{3}(\rho_{x0} + \rho_m),\quad 2\dot H + 3H^2 = 8\pi G \rho_{x_0}.
\end{eqnarray}
It is the same background dynamics of the $\Lambda$CDM model!
At linear perturbative level, we find,
\begin{eqnarray}
\ddot\delta_m + 2\frac{\dot a}{a}\delta_m - 4\pi G\rho_m\delta_m = 0,
\end{eqnarray}
where $\delta_m$ is the density contrast for the pressure less component.
Again, it is the same perturbed equation of the $\Lambda$CDM model!

But now, there is the relation,
\begin{eqnarray}
\delta\rho_x = \frac{\gamma - 1}{2(3 - 2\gamma)}\delta\rho_m.
\end{eqnarray}
Dark energy agglomerates, even it it is in the form of a vacuum energy term! This may modify the $\Lambda$CDM scenario at non-linear level!

\section{Scalar Formulation}

Let us consider  a self-interacting scalar field in the context of the Rastall's theory:
\begin{eqnarray}
 R_{\mu\nu} - \frac{1}{2}g_{\mu\nu}R &=& \phi_{;\mu}\phi_{;\nu} - \frac{2 - \gamma}{2}g_{\mu\nu}\phi_{;\rho}\phi^{;\rho}\nonumber\\
 &+& g_{\mu\nu}(3 - 2\gamma )V(\phi),\\
\Box\phi + (3 - 2\gamma)V_\phi &=& (1 - \gamma)\frac{\phi^{;\rho}\phi^{;\sigma}\phi_{;\rho;\sigma}}{\phi_{;\alpha}\phi^{;\alpha}}.
\end{eqnarray}
This self-interacting scalar field obeys a modified Klein-Gordon equation: it is a non-canonical scalar field.

First, let us inspect the consequences at perturbative level \cite{daoudabis}. Using the newtonian coordinate condition, the perturbed equations when only the "Rastall" scalar field is present, are the following:
\begin{eqnarray}
 \nabla^2\Phi - 3\mathcal{H}\left(\mathcal{H}\Phi + \Phi'\right) + \gamma\left(\mathcal{H}^2 - \mathcal{H}'\right)\Phi =\nonumber\\ 4\pi G\left[\gamma\phi_0'\delta\phi' + (3 - 2\gamma)a^2V_{,\phi}\delta\phi\right]\;,\\
\label{G0i} \mathcal{H}\Phi_{,i} + \Phi'_{,i} = 4\pi G\phi_0'\delta\phi_{,i}\;,\\
\Phi'' + 3\mathcal{H}\Phi' + \left(2\mathcal{H}' + \mathcal{H}^2\right)\Phi + (2 - \gamma)\left(\mathcal{H}^2 - \mathcal{H}'\right)\Phi =\nonumber\\ 4\pi G\left[(2 - \gamma)\phi_0'\delta\phi' - (3 - 2\gamma)a^2V_{,\phi}\delta\phi\right]\;.
\end{eqnarray}

The perturbed equations in the newtonian coordinate condition can be combined to give the following relations:
\begin{eqnarray}
\label{Eq1} & &\Phi'' + 3\mathcal{H}\Phi' + \left(2\mathcal{H}' + \mathcal{H}^2\right)\Phi = \frac{2 - \gamma}{\gamma}\biggr[-k^2\Phi - 3\mathcal{H}
\nonumber\\ & &\left(\mathcal{H}\Phi + \Phi'\right)\biggl] - \frac{2V_{,\phi}a^2}{\gamma\phi_0'}(3 - 2\gamma)\left(\mathcal{H}\Phi + \Phi'\right)\;,\\
\label{Eq2} & &\Phi'' + 3\mathcal{H}\Phi' + \left(2\mathcal{H}' + \mathcal{H}^2\right)\Phi =  - k^2\Phi\nonumber\\ &-& 3\mathcal{H}\frac{2 - \gamma}{\gamma}\left(\mathcal{H}\Phi + \Phi'\right) - \frac{2V_{,\phi}a^2}{\gamma\phi_0'}(3 - 2\gamma)\left(\mathcal{H}\Phi + \Phi'\right)\;.
\end{eqnarray}
These equations can be identical only if $\gamma = 1$.
The other possibility is to suppress any dependence on the spatial coordinates. But, in this case we have just a redefinition of
the background.

The possibility to have real perturbations for $\gamma \neq 1$ can be recovered adding a matter component:
\begin{eqnarray}
\label{G00new} &-& k^2\Phi - 3\mathcal{H}\left(\mathcal{H}\Phi + \Phi'\right) + \gamma\left(\mathcal{H}^2 - \mathcal{H}'\right)\Phi = \nonumber\\
& & 4\pi G a^2\delta\rho_\phi + 4\pi G a^2\delta\rho\;,\\
\label{G0inew} & &k\left(\mathcal{H}\Phi + \Phi'\right) = 4\pi G k \phi_0'\delta\phi + 4\pi G\rho(1 + w)v\;,\\
\label{Gijnew} & &\Phi'' + 3\mathcal{H}\Phi' + 3\mathcal{H}^2\Phi -\gamma\left(\mathcal{H}^2 - \mathcal{H}'\right)\Phi =\nonumber\\
& &4\pi G a^2\delta p_\phi + 4\pi G a^2 \delta p\;.
\end{eqnarray}
This scalar field "requires" matter in order to make sense at perturbative level.

This scalar field formulation using the Rasttal's theory may lead to a new possible unification scenario. For the sound speed we obtain
$c_2^2 = \frac{2 - \gamma}{\gamma}.$
The sound speed is zero if $\gamma = 2$ \cite{gao}. This is in strong contrast with the canonical scalar field, for which the sound speed is always equal to 1.

For $\gamma = 2$, the field equations read,
\begin{eqnarray}
 R_{\mu\nu} - \frac{1}{2}g_{\mu\nu}R &=& \phi_{;\mu}\phi_{;\nu} + g_{\mu\nu}V(\phi),\\
\Box\phi + V_\phi &=& - \frac{\phi^{;\rho}\phi^{;\sigma}\phi_{;\rho;\sigma}}{\phi_{;\alpha}\phi^{;\alpha}}.
\end{eqnarray}
This is equivalent to,
\begin{eqnarray}
R_{\mu\nu} = 8\pi GT_{\mu\nu}.
\end{eqnarray}

The relations for density and pressure are the following:
\begin{eqnarray}
\rho_\phi = \dot\phi^2 + V(\phi), \quad p_\phi &=& - V(\phi).
\end{eqnarray}
Since the sound speed is zero for any type of potential, the formulation presented above opens the possibility for a new unification scenario of dark matter and dark energy as described in reference \cite{scalar}.

\section{Conclusions}

Here, we have exploited two different ways of implementing some quantum effects in classical theory of gravity.
One of them is through external field, a field that is not concerned by the variational principle. We have exemplified this
idea using a scalar-tensor theory of gravity, essentially similar to the Brans-Dicke one, but where the scalar field is not dynamical.
External field may help to implement new cosmological scenarios, combining a GR phase with a BD phase in a high non trivial way.

Later we explored, in a quite phenomenological way, the fact that quantum effects change the energy-momentum tensor conservation law.
If the usual, classical, conservation law is modified, it is possible to keep the advantages of GR and solve some of its
main problems. We have shown explicitly that, in some conditions, it can preserve the advantages of the standard cosmological model, but leading to modifications at non-linear perturbative level, exactly the regime where the standard model has some problems. Possible a new unification framework
for the dark sector of the universe can be implemented using this proposal.

\noindent
{\bf Acknowledgements:} We thank CNPq (Brazil) and FAPES (Brazil) for partial financial support.

\end{document}